# Escaping in couples facilitates evacuation: Experimental study and modeling

Ning Guo [a], Rui Jiang [b], Mao-Bin Hu [a], Jian-Xun Ding [c], Zhong-Jun Ding [c]

[a] *School of Engineering Science, University of Science and Technology of China, Hefei 230026, People's Republic of China*

[b] *MOE Key Laboratory for Urban Transportation Complex Systems Theory and Technology, Beijing Jiaotong University, Beijing 100044, P. R. China*

[c] *School of Transportation Engineering, Hefei University of Technology, Hefei 230009, People's Republic of China*

**Abstract**

In this paper, the impact of escaping in couples on the evacuation dynamics has been investigated via experiments and modeling. Two sets of experiments have been implemented, in which pedestrians are asked to escape either in individual or in couples. The experiments show that escaping in couples can decrease the average evacuation time. Moreover, it is found that the average evacuation time gap is essentially constant, which means that the evacuation speed essentially does not depend on the number of pedestrians that have not yet escaped. To model the evacuation dynamics, an improved social force model has been proposed, in which it is assumed that the driving force of a pedestrian cannot be fulfilled when the composition of physical forces exceeds a threshold because the pedestrian cannot keep his/her body balance under this circumstance. To model the effect of escaping in couples, attraction force has been introduced between the partners. Simulation results are in good agreement with the experimental ones.

## 1. Introduction

Pedestrian evacuation is common in emergency situation threatening life security. If the internal structure of the building is not rational for evacuation process, or pedestrians have not been guided in a right way, the tragedies resulting in death and severe injury are likely to happen. Therefore, it is essential to understand the movement behavior of pedestrians and systematically analyze evacuation characteristic.

To describe and investigate evacuation dynamics, quite a few models have been proposed, which can be classified into macroscopic ones (Hughes, 2002; Huang et al, 2009; Hänseler et al., 2014) and microscopic ones. Microscopic models treat pedestrians as discrete individuals, which can be further classified into two types. In models such as the social force model (Helbing and Molnár, 1995; Helbing et al., 2000; Zeng et al. 2014), the heuristics-based model (Moussaïd et al., 2011), the statistical mechanics-based model (Karamouzas et al., 2014), and steps model (Sivers and Köster, 2015), continuous time and space have been adopted. On the other hand, in models such as the lattice gas model (Muramatsu et al. 1999; Davidich et al. 2013; Flötteröd and Lämmel, 2015) and the floor field model (Burstedde, 2001; Guo et al., 2011; Bandini et al., 2014; Hsu and Chu, 2014), time and space are discretized.

It is generally believed that panic is adverse to evacuation. Helbing et al. (2000) took the panic degree into consideration, and found some evacuation phenomena such as "faster is slower" and "arch". Garcimartín et al. (2014) launched a controlled experiment to demonstrate the "faster is slower" effect. By allowing pushing one another or not, competitive or non-competitive egress condition was proposed. It was found that competitive egress produced 2s longer evacuation time in experiments of 85 pedestrians.

Evacuation after earthquake or fire is often accompanied by darkness or smoke. Therefore visibility is an important factor in escape dynamics. Isobe et al. (2004) studied evacuation process from a room without visibility, finding that pedestrian behaviors consist of biased random walk to the wall and moving along the wall. On this basis, Nagatani et al. (2004) explored step number of biased random walk and the first touch point along the wall. Guo et al. (2012) investigated the pedestrian evacuation with no visibility through simulation and experiment, and found that pedestrians prefer to follow others or move along walls and obstacles. Yue et al. (2010) simulated pedestrian dynamics with impaired vision. They showed that the evacuation time depends not only on the sight radius but also on the width and utilization rate of exit. Ma et al (2013) introduced view radius and the cluster effect in the simulation, and showed that short view radius could even lead to failure to escape. Yuan et al. (2009) investigated the effects of visibility range and guiders in evacuation, and also found that small visibility range was adverse to evacuation. Hou et al. (2014) investigated the effect of trained leaders on evacuation dynamics with limited visibility. They showed that when the number of leaders is equal to the number of exits and each leader heads to a different exit, the evacuation is more efficient.

Route choices for escaping a building also affect the evacuation efficiency. Zhao and Gao (2010) formulated the route choice of pedestrians to multiple exits, considering distance potential and congestion level near exits, and found that pedestrians distribute heterogeneously under the premise. Alizadeh (2011) used a class of lattice potentials weighting route distance, and found human psychology and pedestrian distribution play important roles in the evacuation. Kretz (2009) analyzed the influences of route distance congestion on route choice using the space potential, showing pedestrians are liable to minimize the travel time. Guo and Huang (2011) proposed a method to compute lattice potential measuring factors consisting of route distance, congestion and route capacity. Simulations illustrated that excessive or limited sensitivity of pedestrians to the route capacity may be unhelpful.

Obstacle installment near the exit can either facilitate or impede evacuation process. Kirchner et al. (2003) utilized a modified cellular automaton model to simulate the situation in which an obstacle has been put in front of the exit. It was found that evacuation time has decreased. Yanagisawa et al. (2009) showed that obstacles located symmetrically near the exit became harmful for evacuation dynamics, and therefore proposed that the optimization approach is to shift the obstacle slightly away from the exit center. Frank et al. (2011) studied the influence of diverse distance between panel-like obstacle and exit, and showed that an obstacle could not guarantee better chance of survival. Jiang et al. (2014) carried out an experiment to validate simulated optimal obstacle layout, demonstrating that two pillars on both sides of the exit minimized the escape time. The evacuation time of 80 pedestrians decreases from 29.5s to 27s. Oh et al. (2014) studied escape process in the condition of various exit angles via simulation and mouse experiment, and claimed that the angle for shortest evacuation time was round 60 degrees.

Grouping, such as families and friends, is a common phenomenon in pedestrian crowds. Moussaïd et al. (2010) added vision forces, attractive forces and repulsive forces in groups in the social force model, and reproduced the V-like pattern in 3-member groups. Qiu et al. (2010) proposed a model considering the influence of the group. Simulation shows that the structure, interaction of groups and scale of group members have main impact over the crowd. Xu et al. (2010) proposed bonding effects among group members, and found that social relationship is negative to walking speed. Koster et al. (2011) developed a lattice gas model, and confirmed that the larger the group size, the slower the average speed.

To our knowledge, the effect of grouping on the evacuation dynamics has not been studied. In this paper, we carry out experiments to study the impact of escaping in couples. It has been found that escaping in couples can decrease the average evacuation time than escaping in individual. To model the evacuation dynamics, an improved social force model has been proposed, in which it is assumed that the driving force of a pedestrian cannot be fulfilled when the composition of physical forces exceeds a threshold because the pedestrian cannot keep his/her balance under this circumstance. The improved model is shown to be able to correctly reproduce the evacuation process while the original social force model fails. To model the effect of escaping in couples, attraction force has been introduced to describe the interaction between partners. Simulation results are in good agreement with the experimental ones.

The remainder of this paper is organized as follows. Section 2 describes the experimental setup and results. Section 3 presents models and simulation results. Section 4 concludes the paper.

## 2. Experiment

2.1 Experimental setup

The experiments were performed in November 16, 2014 in Hefei University of Technology. Experiments were conducted in an artificial room with one exit. The size of the virtual room was 7*7 $m^2$, at the boundary of which some chairs and two rostrum tables were placed as walls. The exit has a width of 0.8 m, and is located in the centre left of one wall, see Fig.1. A total of 50 participants (graduate students, 39 male, 11 female) took part in the experiments, and they were naïve to the purpose of experiments.

Participants were asked to escape from the room as soon as possible. Two sets of experiments have been implemented. In the first set of experiments, 50 participants were requested to regard other participants as strangers and escape individually, see Fig. 1(a). They were distributed randomly at the beginning. In the second set of experiments, 50 participants were divided into 25 couples. Each participant and his/her partner wore the same hats, but different couples had different hats in color and/or style, see Fig. 1(b). Participants are requested to imagine their partners as family members or friends. Before the evacuation, participants in a couple stood near each other, and the couples were distributed randomly.

The two sets of experiments were conducted alternately with 6 replications respectively. The motion of each participant was recorded by video camera (SONY HDR-CX510E), and the evacuation times were recorded manually.

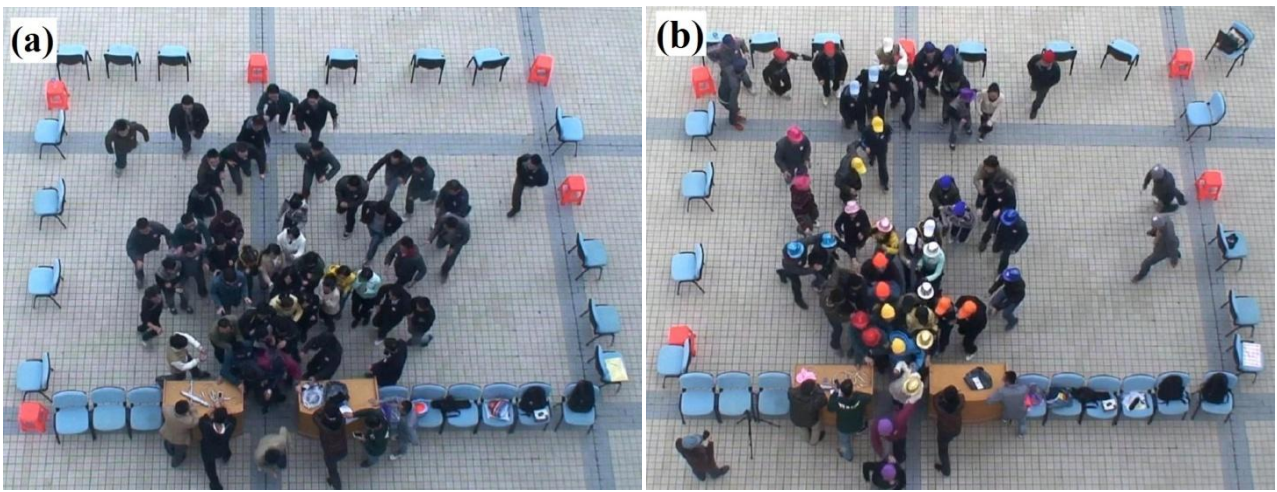

Fig. 1  Snapshots of experiments. (a) escape in individual and (b) escape in couples.

Table 1 Evacuation time (unit: second) in the two sets of experiments.

|  | Experiment | | | | | | | | Simulation | |
| --- | --- | --- | --- | --- | --- | --- | --- | --- | --- | --- |
|  | No. 1 | 2 | 3 | 4 | 5 | 6 | Average time | Standard deviation | Average time | Standard deviation |
| First set | 16.96 | 17.97 | 16.70 | 17.72 | 17.50 | 17.64 | 17.41 | 0.44 | 17.48 | 0.67 |
| Second set | 18.44 | 16.02 | 16.00 | 16.01 | 15.80 | 16.78 | 16.51 | 1.00 | 16.40 | 1.08 |

2.2 Experimental results

The experiments reveal that escaping in couples does have effect on evacuation process. Table. 1 shows the evacuation time in the two sets of experiments. The average evacuation time in the first set of experiments is 17.41 s, and the standard deviation is 0.44 s. On the contrast, the second set of experiments has shorter average evacuation time (16.51 s), but the standard deviation (1.00 s) is larger.

Fig. 2(a) exhibits the leaving time of each pedestrian. One can see that time gap between two consecutive pedestrians varies in the evacuation process. Table. 2 shows the maximum and minimum number of escaped pedestrians per second. In both sets of experiments, the average maximum number is 5.2 pedestrians. However, the average minimum number is 1.2 pedestrians when pedestrians escape individually, which is smaller than 1.5 pedestrians when pedestrians escape in couples.

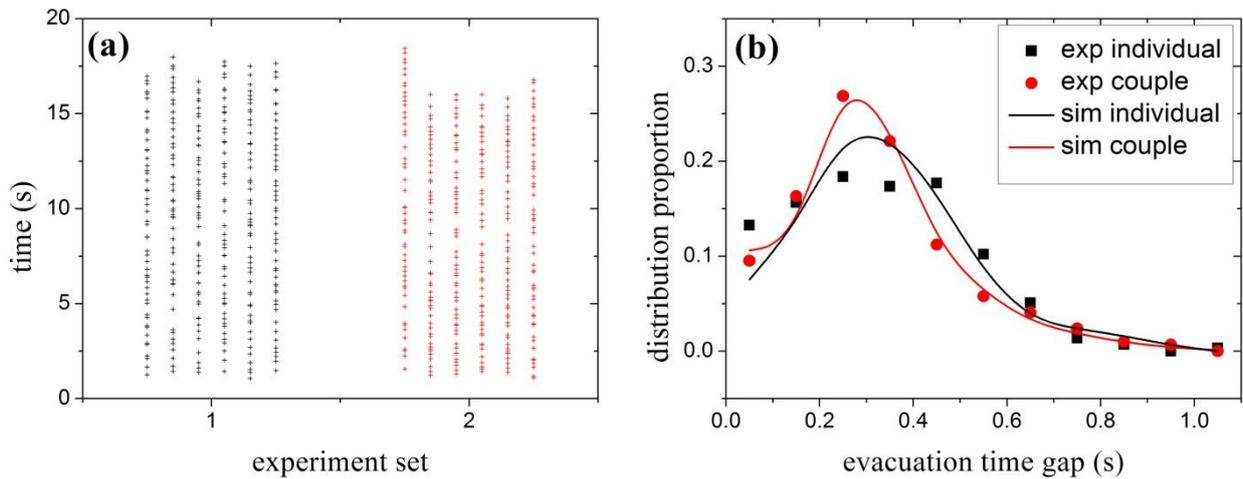

Fig. 2 Evacuation characteristics. (a) leaving time of each pedestrian. (b) evacuation time gap distribution.

Table 2 Maximum and minimum number of escaped pedestrians per second.

| first set | No. 1 | 2 | 3 | 4 | 5 | 6 | experiment average | simulation average |
|---|---|---|---|---|---|---|---|---|
| max | 5 | 5 | 5 | 6 | 5 | 5 | 5.2 | 5 |
| min | 1 | 1 | 1 | 1 | 2 | 1 | 1.2 | 1.4 |
| second set | | | | | | | | |
| max | 6 | 4 | 5 | 5 | 5 | 6 | 5.2 | 5.2 |
| min | 1 | 2 | 2 | 2 | 1 | 1 | 1.5 | 1.2 |

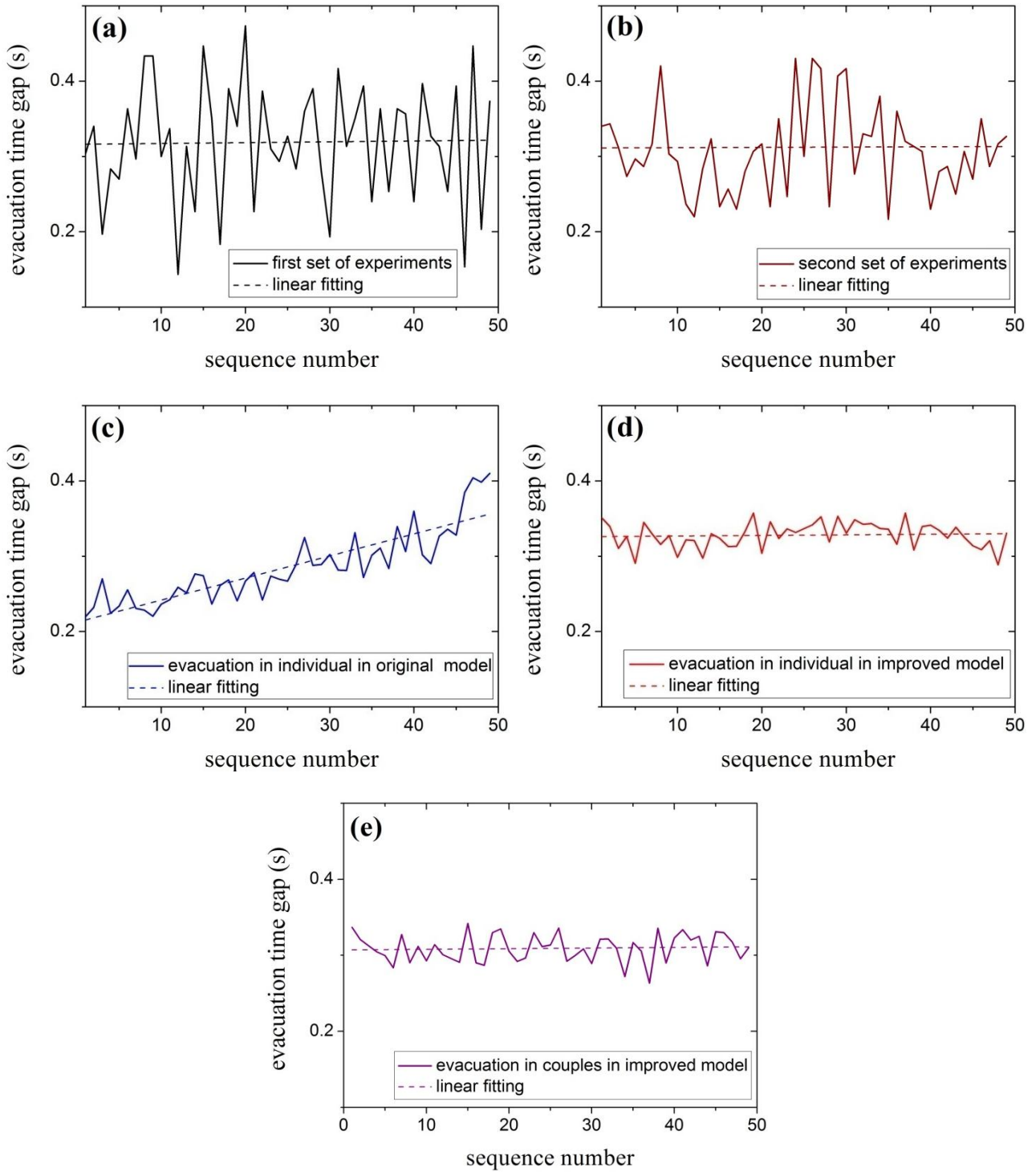

Fig. 3 Average evacuation time gap between consecutive pedestrians in (a) first set and (b) second set of experiments. (c) shows simulation results of escaping in individual in the original social force model, (d) and (e) show simulation results of the improved social force model. The experimental results are averaged over 6 replications and the simulation results are averaged over 50 replications. The dashed lines are linear fitting results and are guides for the eyes.

Fig. 2(b) shows the distribution of evacuation time gap. It can be seen that the distribution is narrower and has a larger peak value in the second set of experiments. As a result, the standard deviation of the evacuation time gap is 0.1804 s when pedestrians escape in couples, which is smaller than 0.1882 s when pedestrians escape individually. This is a little surprising because the standard deviation of evacuation time is larger when pedestrians escape in couples. This result might be related to the formation and the lifetime of arches. It is likely that when pedestrians escape in couples, the formation probability of arches decreases. This explains the short evacuation time in replications 2-6 in the second set of experiments. Therefore, both the average and the standard deviation of evacuation time gap decrease. However, once an arch has formed, it can survive a longer time. This will induce more large evacuation time gaps, which explains the long evacuation time in replication 1 in the second set of experiments. As a result, the standard deviation of evacuation time increases.

Fig. 3(a) and 3(b) show the average evacuation time gap between the *i* th escaped pedestrian and the *i+1* th escaped pedestrian. Despite of the fluctuations, which are expected to be suppressed with increase of experiment replications, the time gap is essentially constant in both sets of experiments. This implies that the evacuation speed essentially does not depend on the number of pedestrians that have not yet escaped.

### 3. Model and simulation

3.1 Social force model

Firstly, we simulate the evacuation dynamics of the 1$^{st}$ set of experiments by using the original social force model. In the social force model, the motion of a pedestrian is described by three different components: driving force, repulsive force between pedestrians, and repulsive force between pedestrian and wall. The equation of motion is as follows,

$$m_i \frac{d\vec{v}_i}{dt} = \vec{f}_i^0 + \sum_j \vec{f}_{ij} + \sum_W \vec{f}_i^{wall} \quad (1)$$

Here $m_i$ is mass of pedestrian *i*. The driving force $\vec{f}_i^0$ represents the pedestrian's motivation to walk in a given direction with desired speed, which reads

$$\vec{f}_i^0 = m_i \frac{v_i^0 \vec{e}_i^0 - \vec{v}_i(t)}{\tau} \quad (2)$$

Here $v_i^0$ and $\vec{e}_i^0$ are respectively magnitude and direction of the pedestrian's desired speed. $\vec{v}_i(t)$ is the current speed, and $\tau$ denotes the relaxation time.

The repulsive force $\vec{f}_{ij}$ describes pedestrian *i*'s interaction with other pedestrians, which includes socio-psychological force to stay away from others and physical contact force.

$$\vec{f}_{ij} = A \cdot \exp[(r_{ij} - d_{ij})/B]\vec{n}_{ij} + m_i \cdot g(r_{ij} - d_{ij})(k\vec{n}_{ij} + \kappa \Delta v_{ji}^t \vec{t}_{ij}) \tag{3}$$

$$g(x) = \begin{cases} x, & x > 0 \\ 0, & x \leq 0 \end{cases} \tag{4}$$

In Eq. (3), the first term on the right hand side denotes the socio-psychological force, and the second term denotes the physical contact force. Pedestrians *i* and *j* are regarded as cylinders with radius $r_i$ and $r_j$, respectively. $r_{ij}$ denotes the sum of the radius of the two pedestrians, namely $r_{ij} = r_i + r_j$. $d_{ij}$ is the distance between the centers of the two pedestrians. $\vec{n}_{ij}$ represents the unit normal vector from pedestrian *j* to *i*, $\vec{t}_{ij}$ means the unit tangential vector,

$$\vec{n}_{ij} = (\frac{x_i - x_j}{d_{ij}}, \frac{y_i - y_j}{d_{ij}}) \tag{5}$$

$$\vec{t}_{ij} = (-\frac{y_i - y_j}{d_{ij}}, \frac{x_i - x_j}{d_{ij}}) \tag{6}$$

where $x_i, x_j$ and $y_i, y_j$ are respectively the horizontal and vertical coordinates of the center of the two pedestrians. $\Delta v_{ji}^t$ means the tangential velocity,

$$\Delta v_{ji}^t = (\vec{v}_j - \vec{v}_i) \cdot \vec{t}_{ij} \tag{7}$$

where $k\vec{n}_{ij}$ and $\kappa \Delta v_{ji}^t \vec{t}_{ij}$ are body pressure and friction force respectively. *A*, *B*, *k* and *κ* are four parameters.

The repulsive force $\vec{f}_i^{wall}$ from the wall is modeled analogously,

$$\vec{f}_i^{wall} = A \cdot \exp[(r_i - d_i^{wall})/B]\vec{n}_i^{wall} + m_i \cdot g(r_i - d_i^{wall})(k\vec{n}_i^{wall} + \kappa \Delta v_i^t \vec{t}_i^{wall}). \tag{8}$$

Here, $d_i^{wall}$ means vertical distance to the wall, $\vec{n}_i^{wall}$ denotes the unit normal vector from the wall to pedestrian *i*, and $\vec{t}_i^{wall}$ is the unit vector tangential to the wall. The tangential velocity $\Delta v_i^t = -\vec{v}_i \cdot \vec{t}_i^{wall}$, since speed of the wall is zero.

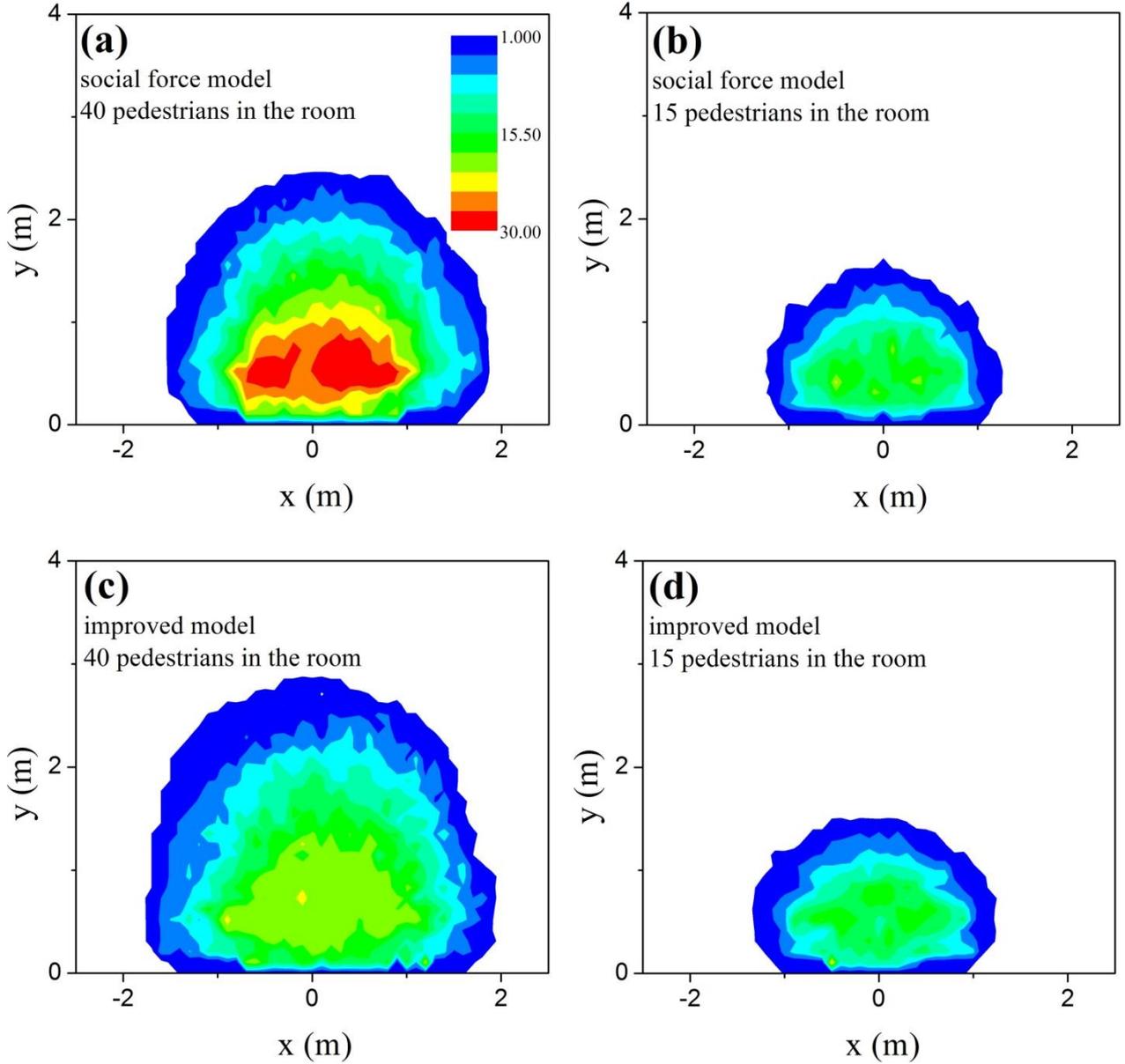

Fig. 4 Simulation results of distribution of the pressure-like parameter $P$. Initially there are 50 pedestrians in the room, and they escape individually. (a) and (b) are simulation results of the original social force model. (c) and (d) are simulation results of the improved model.

In the simulation, the parameters are set as follows: $m_i = 70 kg$, $v_i^0 = 1.5 m/s$, $r_i = 0.225m$, $\tau = 0.5s$, $k = 1200 s^{-2}$. We set $\kappa = 0$, because the evacuation experiment is just like a fire drill without real panic atmosphere. We also set $A = 0$, i.e., we neglect the socio-psychological repulsive forces, because pedestrians do not care to stay away from others in the evacuation process.

With these parameters, the average evacuation time is 17.44 s, which is in good agreement with the experimental result. However, as can be seen from Fig. 3(c), instead of being essentially constant, the evacuation time gap gradually increases. This means that with more pedestrians escaped from the room, the evacuation speed lowers down. This is due to the accumulation effect of driving force. To quantify the effect, we define a pressure-like parameter $P$ as the magnitude of repulsive physical contact forces between pedestrian $i$ and other pedestrians.

$$P_i = \frac{1}{m_i} \sum_{j=1, j \neq i}^{N} |\overrightarrow{f_{ij}}| \qquad (9)$$

where $N$ is the number of pedestrians still not yet escaped.

Fig. 4(a) and (b) show the distribution of $P$, which is averaged over 500 replications. One can see that roughly speaking, the closer to the door, the larger $P$ is. Moreover, with more pedestrians escaped from the room (Fig. 4(b)), the magnitude of $P$ in the vicinity of the door remarkably decreases. When $P$ is large, pedestrians near the door will be pushed by large force. Therefore, they escape quickly. With the decrease of $P$, the pushing force decreases, which lowers down evacuation speed. This simulation result is analogous to the fluid dynamic problem. Suppose there is a barrel of beer and there is a tap at the bottom of the barrel. The more beer has been drunk, the smaller the flow speed of the beer from the tap, because the pressure near the tap lowers down. Here the driving force is analogous to the gravity and the pressure-like parameter $P$ is analogous to the pressure of beer in the barrel.

3.2 Improved social force model

Now we propose an improved social force model to overcome the deficiency presented above. When squeezed in the crowd, a pedestrian can be pushed away if the composition of contact forces from other pedestrians and walls is large. Under this circumstance, the pedestrian would firstly try to keep body balance and is thus unable to fulfill the driving force temporarily. Based on this fact, the driving force is modified

$$\overrightarrow{f_i^0} = \begin{cases} m_i \dfrac{v_i^0 \overrightarrow{e_i^0} - \overrightarrow{v_i(t)}}{\tau}, & f_i^{contact} \leq f_c \\ 0, & f_i^{contact} > f_c \end{cases} \qquad (10)$$

Here $f_i^{contact} = \left| \sum_j \overrightarrow{f_{ij}} + \sum_W \overrightarrow{f_i^{wall}} \right|$ means the composition of physical forces, and $f_c$ is a threshold parameter.

Next we present simulation results of the improved social force model, in which $f_c = 10 m_i$, $v_i^0 = 1.8 m/s$ and other parameters are same as before. As can be seen from Table 1, the average evacuation time is 17.48 s, and the standard deviation is 0.67 s, which are in pretty good agreement with the experimental results. Table 2 shows that simulation results of the average maximum and minimum number of escaped pedestrians per second are respectively, 5 and 1.4 pedestrians, which are also in agreement with the experimental ones. Fig. 2(b) shows that simulation results of the distribution of evacuation time gap are in consistent with the experimental ones.

More importantly, Fig. 3(d) shows that the evacuation time gap is essentially a constant as observed in the experiments. The distribution of the pressure-like parameter $P$ has been shown in Fig. 4(c) and (d), which is remarkably different from that in the original model. One can see that magnitude of $P$ in the vicinity of the door is essentially independent of the number of pedestrians that have not yet escaped. This is because, as shown in Fig. 5, the more pedestrians in the room, the more pedestrians not able to fulfill the driving force. Consequently, the evacuation speed essentially does not change.

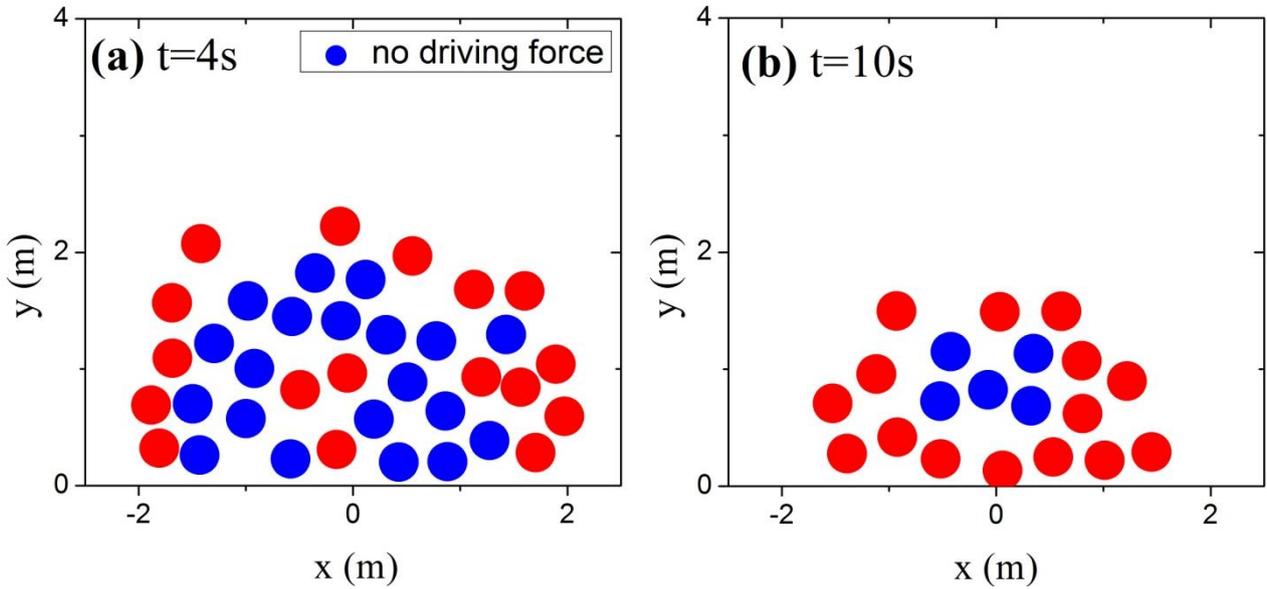

Fig. 5  Two snapshots in the evacuation process in one replication of the simulation. Initially there are 50 pedestrians in the room, and they escape individually. The red (blue) dots mean that pedestrians are able (unable) to fulfill the driving force at the moment.

3.3 Attractive force between partners

An attraction force between the partners is introduced to depict the effect of escaping in couples in evacuation process. The attraction force exerted from pedestrian $h$ on his/her partner $i$ is proposed as follow,

$$\overrightarrow{f_{ih}^{group}} = \begin{cases} m_i \cdot (C_1 - C_1 \cdot e^{\frac{r_{ih} - d_{ih}^{group}}{D}}) \cdot \overrightarrow{n_{ih}^{group}}, & f_i^{contact} \leq f_c \quad and \quad r_{ih} \leq d_{ih}^{group} \quad and \quad d_i^{exit} \geq d_h^{exit} \\ m_i \cdot (C_2 - C_2 \cdot e^{\frac{r_{ih} - d_{ih}^{group}}{D}}) \cdot \overrightarrow{n_{ih}^{group}}, & f_i^{contact} \leq f_c \quad and \quad r_{ih} \leq d_{ih}^{group} \quad and \quad d_i^{exit} < d_h^{exit} \\ 0, & f_i^{contact} > f_c \quad or \quad r_{ih} > d_{ih}^{group} \end{cases} \quad (13)$$

Here $d_{ih}^{group}$ represents the distance between pedestrians $i$ and $h$. $r_{ih}$ denotes the sum of the radius of the two pedestrians. $\overrightarrow{n_{ih}^{group}} = (x_h - x_i, y_h - y_i)/|(x_h - x_i, y_h - y_i)|$ means the unit direction vector of the attraction force. $d_i^{exit}$ denotes the distance from pedestrian $i$ to the center of exit. $D$ is a parameter. When pedestrian $h$ is closer to the door, coefficient $C_1$ is used in the attraction force. In contrast, when pedestrian $i$ is closer to the door, a different coefficient $C_2$ is used. The two coefficients need to be calibrated, which may be either equal or unequal to each other. Since the attraction force is also socio-psychological force, pedestrians cannot fulfill the force either, when the composition of physical forces exceeds the threshold $f_c$. Finally, we assume that the attraction force works only when the two pedestrians are separated ($r_{ih} \leq d_{ih}^{group}$).

Now we present simulation results of evacuation process when pedestrians escape in couples. The parameters are set as $C_1 = 2$ m/s$^2$, $C_2 = 1$ m/s$^2$, $D = 0.1$ m, other parameters are same as before.

Table 1 shows that comparing with escaping in individual, the average evacuation time decreases to 16.40 s, and the standard deviation increases to 1.08 s. This is in consistent with the experimental finding. Table 2 shows that the average maximum and minimum number of escaped pedestrians per second are respectively, 5.2 and 1.2 pedestrians, still in agreement with the experimental ones. Fig. 2(b) shows the distribution of evacuation time gap (red solid line), which is narrower and has large peak value. As a result, the standard deviation of the evacuation time gap is 0.1735 s, which is smaller than 0.1816 s (corresponding to the black solid line). This is also in consistent with the experiments. Moreover, Fig. 3(e) shows that the average evacuation time gap is essentially constant as observed in the experiments.

## 4. Conclusion

In this paper, we have experimentally studied the impact of escaping in couples on the evacuation dynamics. Two sets of experiments have been implemented, in which pedestrians are asked to escape either in individual or in couples. The experiments show that escaping in couples can decrease the average evacuation time of 50 pedestrians from 17.41s to 16.51s. Another important finding is that the average evacuation time gap is essentially constant, which means that the evacuation speed essentially does not depend on the number of pedestrians that have not yet escaped.

We have carried out simulation of the evacuation process by using the social force model. It has been shown that the model fails to reproduce the constant average evacuation time gap due to accumulation effect of driving force. We have proposed an improved model to overcome the deficiency in the original social force model, in which it is assumed that the driving force of a pedestrian cannot be fulfilled when the composition of physical forces exceeds a threshold because the pedestrian cannot keep his/her balance under this circumstance. To model the effect of escaping in couples, attraction force has been introduced between the partners. Simulation results are in good agreement with the experimental ones.

In our future work, more experiments should be carried out to examine the results reported in this paper. Moreover, the impact of escaping in triples or in more large groups also needs to be studied.

Finally, we would like to mention that one should be very careful to generalize the conclusion that escaping in couples always facilitates evacuation in real emergency situation. This is because the experiment is like a fire drill without panic atmosphere. In real emergency situation with a matter of life and death, even if our model is applicable, the parameters are likely to be different. Suppose some parameters have changed, e.g., $\kappa = 8$ $m^{-1}s^{-1}$. With all other parameters as in Fig. 3(e), simulation shows that comparing with escaping in individual, escaping in couples increases the average evacuation time from 16.40 s to 27.75 s. Therefore, empirical study is an indispensable task for the practical application.